\documentclass[12pt]{article}

\usepackage{newtxtext,newtxmath}
\usepackage{bm}

\usepackage{graphicx}

\usepackage[letterpaper,margin=1in]{geometry}

\renewenvironment{abstract}
	{\quotation}
	{\endquotation}

\date{}

\makeatletter
\renewcommand{\fnum@figure}{\textbf{Figure \thefigure}}
\renewcommand{\fnum@table}{\textbf{Table \thetable}}
\makeatother

\usepackage{SciAdv_Temp/scicite}

\usepackage{url}

\def\scititle{
	Strain control of mid-IR spectroscopic nonlinear photocurrents in PtSe$_2$
}
\title{\bfseries \boldmath \scititle}

\author{
	M. Gerlei$^{1\ast}$,
	G. J. de Coster$^{2,3}$,
	J. Papp$^{1}$,
    S. Heiserer$^{1}$,\and
    S. Schlosser$^{1}$,
    G. S. Duesberg$^{1}$,
    P. Seifert$^{1\ast}$\\
	\small$^{1}$Institute of Physics \& Center for Integrated Sensor Systems (SENS),\\
    \small University of the Bundeswehr Munich, Neubiberg, Germany\and
	\small$^{2}$DEVCOM Army Research Laboratory, Adelphi, MD, USA\and
    \small$^{3}$MIT Institute for Soldier Nanotechnologies, Cambridge, MA, USA\and
	\small$^\ast$Corresponding author. Email: martin.gerlei@unibw.de, paul.seifert@unibw.de
}

\begin{document} 

\maketitle

\begin{abstract} \bfseries \boldmath
Semi-metallic noble-metal dichalcogenides are promising materials for infrared optoelectronics, yet the origin and tunability of their nonlinear optical responses remain poorly understood. Here, we demonstrate \textit{in-situ} mechanical control of photon-drag photocurrents in polycrystalline PtSe$_2$ thin films grown on flexible polyimide substrates. Using polarization-resolved mid-infrared photocurrent spectroscopy under uniaxial tensile strain, we observe pronounced strain-dependent changes in resistivity, photoconductivity, and helicity-dependent nonlinear photocurrents. The spectral response is consistent with optical excitations across the spin-orbit-coupling-induced (SOC) band gap near the K point. We develop a theoretical framework attributing the photocurrent primarily to photon-drag induced by photon-momentum symmetry breaking. Strain modifies its magnitude and polarization dependence through deformation potentials that tune the SOC-induced band gap. These results establish mechanical strain as a route toward reconfigurable infrared polarization detection.
\end{abstract}

\section*{INTRODUCTION}
\noindent Two-dimensional (2D) transition metal dichalcogenides (TMDs) have emerged as a versatile platform for exploring novel physics in materials, and developing next-generation optoelectronic and sensing technologies \cite{lemme_nanoelectromechanical_2020, Bhimanapati2015, Ponraj2016, Cheng2019, MakAndHeinz2010, Lee2013}.
Recently, platinum diselenide (PtSe$_2$) from the less explored family of noble-metal dichalcogenides (NMDCs) has attracted increasing attention:
its tunable electronic structure, mechanical flexibility, chemical and thermal stability in air give it potential for integration into next-generation optoelectronic and sensing devices. 
Importantly, PtSe$_2$ can be synthesized using scalable techniques such as thermally assisted conversion (TAC) of pre-deposited platinum films (Fig.~\ref{fig:fig1}a), chemical vapor deposition (CVD), and molecular beam epitaxy (MBE), all of which are compatible with wafer-scale processing and device integration \cite{OBrien2016, Yim2016, Chen2020, Prechtl2021, Prechtl2025, HilseMBE, Pierre2023}.
Moreover, its pronounced piezoresistivity is a focus of interest for the development of pressure sensors, microphones and strain-tunable electronic devices \cite{Wagner2018, Boland2019, Heiserer2025, Lukas2025}.

In its bulk form, PtSe$_2$ exhibits a semi-metallic behavior, but when thinned to a few atomic layers, it undergoes a transition to a semiconducting state with a finite band gap \cite{Wang2015, Ansari2019}.
The optical absorption of PtSe$_2$ extends from visible in the monolayer limit to mid-infrared in the bulk, due to its spin-orbit coupling (SOC) induced low-energy band gap at the $K$-point (Fig.~\ref{fig:fig1}b) \cite{Yim2016, Yim2018, Prechtl2021, Han2019}.
Moreover, recent theoretical and experimental studies have revealed that PtSe$_2$ hosts exotic electronic states, including type-II Dirac fermions and topologically protected surface bands \cite{Wang2015}.
These combined characteristics thus make PtSe$_2$ a compelling material for robust and sensitive infrared (IR) photodetection technologies \cite{Lan2020}.
In its pristine form, PtSe$_2$ is expected to exhibit isotropic behavior due to its crystallization in the 1T phase with hexagonal symmetry (Fig.~\ref{fig:fig1}c).
However, experimental observations of polarization-sensitive photoresponses, such as circular dichroism, second harmonic generation or helicity dependent photocurrents suggest the presence of intrinsic or extrinsic symmetry breaking mechanisms in NMDCs, particularly prominent in the mid-IR to THz range \cite{Hemmat2023, Ganichev, Chen2024, Hemmat2022, Chen2025, Yadav2024, Guo2020, DeCoster2026, Cheng2023}.
In polycrystalline films, anisotropies could potentially arise from factors such as grain orientation in growth direction, defect gradients, interfacial strain, or external mechanical deformation \cite{Chen2024, Liu2025}.
Understanding and controlling these effects are crucial to exploiting PtSe$_2$ in reconfigurable and polarization-sensitive optoelectronic devices.

Here, we investigate the interplay between mechanical deformation and mid-IR photoresponse of polycrystalline PtSe$_2$ thin films synthesized directly on flexible polyimide substrate using a low-temperature synthesis process \cite{Yim2016, Boland2019}.
We built an experimental setup, depicted in Fig.~\ref{fig:fig1}e inset and in Fig. S1 that allowed us to uniaxially strain the film while conducting polarization-resolved spectroscopic IR measurements.

The experiments reveal mechanical control of resistivity, photoconductivity and helicity-dependent nonlinear optical (NLO) photoresponses.
Further, we establish a theoretical framework in which our observation of mid-IR generated photocurrent is best described by a photon drag-effect (PDE) involving excitations at the $K$-point of bulk PtSe$_2$, modified by deformation potentials controlling the SOC band gap  \cite{Nagaosa,Parker2019,Bir1974}.

\section*{RESULTS}
\subsection*{Direct growth of semi-metallic \protect\NoCaseChange{PtSe$_2$} on a strainable substrate}
\label{results:1}

\begin{figure} 
	\centering
	\includegraphics[width=1\textwidth]{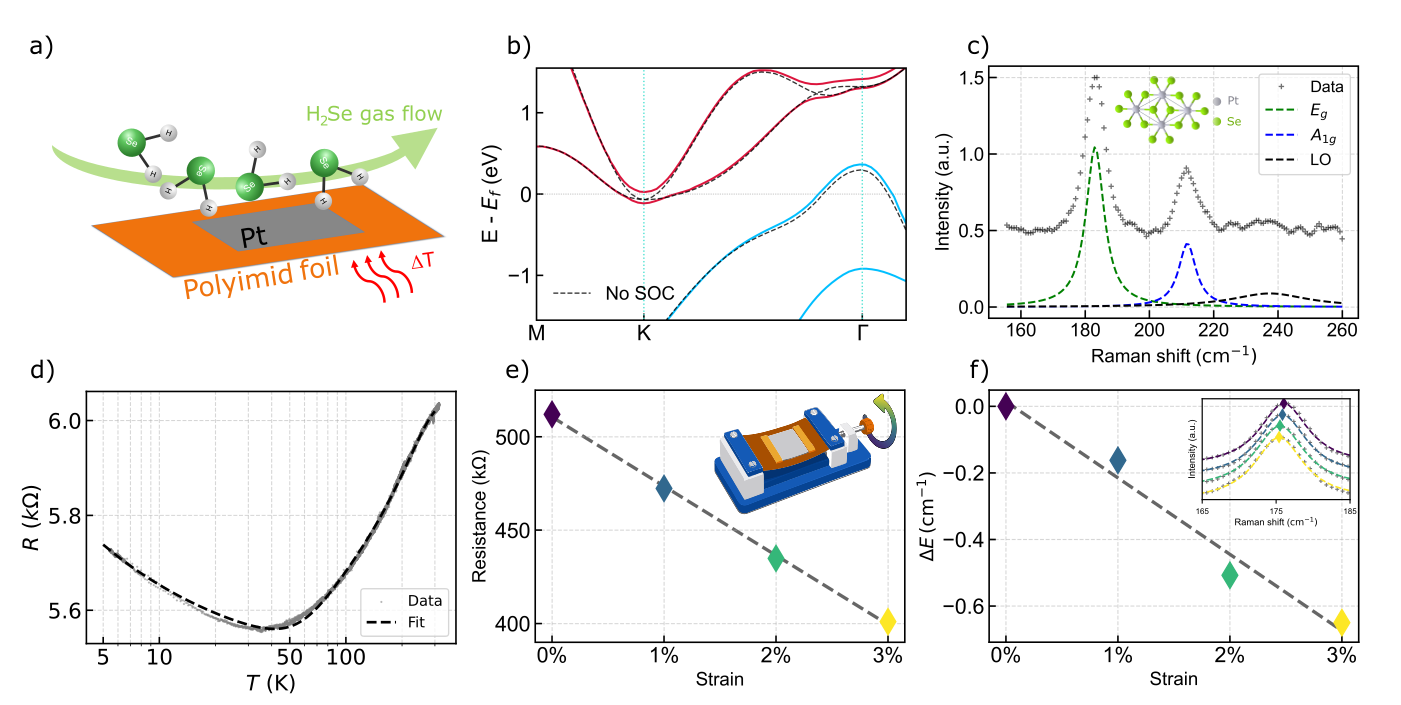} 

	\caption {\textbf{Strainale semimetallic PtSe$_2$ device} - a) Low temperature thermally assisted conversion (TAC) of thin platinum layers enables the synthesis of polycrystalline multilayer PtSe$_2$ films directly on non-conventional, flexible substrates such as polyimide foil. b) Schematic band structure of bulk PtSe$_2$ showing spin-orbit coupling induced gap at the $K$-point. c) Characteristic Raman spectrum of 1T PtSe$_2$ on polyimide with Lorentzian fits to the in plane $E_g$ and out of plane $A_{1g}$ mode as well as longitudinal optical ($LO$) modes. The inset depicts the hexagonal 1T PtSe$_2$ crystal structure. d) Temperature dependent electrical transport measurement of TAC synthesized PtSe$_2$ film on Si/SiO$_2$. The dashed line is the corresponding fit containing activation energy. e) Resistance of polycrystalline PtSe$_2$ on polyimide as a function of uniaxial tensile strain, showing negative piezoresistivity. The inset depicts a schematic of the experimental setup. f) Energy of the in-plane ($E_g$) Raman mode red-shifts as a function of uniaxial tensile strain.}
    \label{fig:fig1}
\end{figure}

\noindent Polycrystalline multilayer platinum diselenide (PtSe$_2$) films were obtained via TAC of predeposited platinum. 
The selenisation was performed at 400~°C in order to achieve direct growth on a flexible substrate, in this case polyimide foil (Fig.~\ref{fig:fig1}a).
This ensures that the PtSe$_2$ is in immediate mechanical contact with the underlying substrate and the contact pads.

The resulting PtSe$_2$ predominantly crystallizes in the very stable hexagonal 1T phase, as verified via Raman spectroscopy (Fig.~\ref{fig:fig1}c).
At its target thickness of 10~nm, the PtSe$_2$ film is expected to show bulk semi-metallic behavior with p-type band at the $\Gamma$-point and a pair of SOC-split bands at the $K$-point with a band splitting of around 
160 meV (Fig.~\ref{fig:fig1}b).

Films on the order of 10~nm thickness show a resistance that increases with temperature, consistent with the semi-metallic nature of bulk PtSe$_2$.
The transport data in Figure~\ref{fig:fig1}c can be fitted with a resistance versus temperature model for parallel semiconducting and metallic channels that captures the crossover from metallic to semiconducting behavior as temperature increases.
With
$R = (\frac{1}{R_\infty e^{\Delta/k_{B}T}} + \frac{1}{R_{e-p}(T)})^{-1} $
we obtain an activation energy of $\Delta = 153$ meV, which is in good agreement with the SOC band gap (Fig.~\ref{fig:fig1}d).

Electrical measurements under uniaxial strain revealed a negative piezoresistive effect, where resistance decreases with increasing strain consistent with previous observations in literature \cite{Wagner2018, Boland2019, Heiserer2025, Lukas2025}.
Current–voltage characteristics measured at 0–3\% strain show systematic resistance reduction, yielding a gauge factor of $G_f~\approx~-7$ (Fig~\ref{fig:fig1}e).
The observed negative piezoresistivity indicates that the strain-induced resistance change cannot be attributed solely to geometric effects.
Instead, recent work indicates that mechanical deformation modifies the electronic band structure and the density of states near the Fermi level, increasing both intra-grain conductivity, as well as inter-grain hopping in polycrystalline PtSe$_2$ \cite{Heiserer2025}.

Simultaneously, we identify a red-shift of the $E_g$ Raman mode under strain, which indicates strain induced phonon softening.
The increased phonon wavelength corresponds to a reduction in the vibrational energy, reflecting the expansion of the lattice constant under tensile deformation (Fig.~\ref{fig:fig1}f).
This observation shows that the strain applied to the polyimide substrate is directly transferred to the PtSe$_2$ film uniformly.
In contrast, in previous work on strain induced Raman shifts in free standing PtSe$_2$, the strain was applied directly to the inhomogeneous PtSe$_2$ film, leading to a highly non-uniform strain distribution with a geometrically induced blue-shift of the $E_g$ mode \cite{Heiserer2025}.

\subsection*{Mid-IR optoelectronic response from the $K$-point gap}
\label{results:2}

\begin{figure}
	\centering
	\includegraphics[width=1\textwidth]{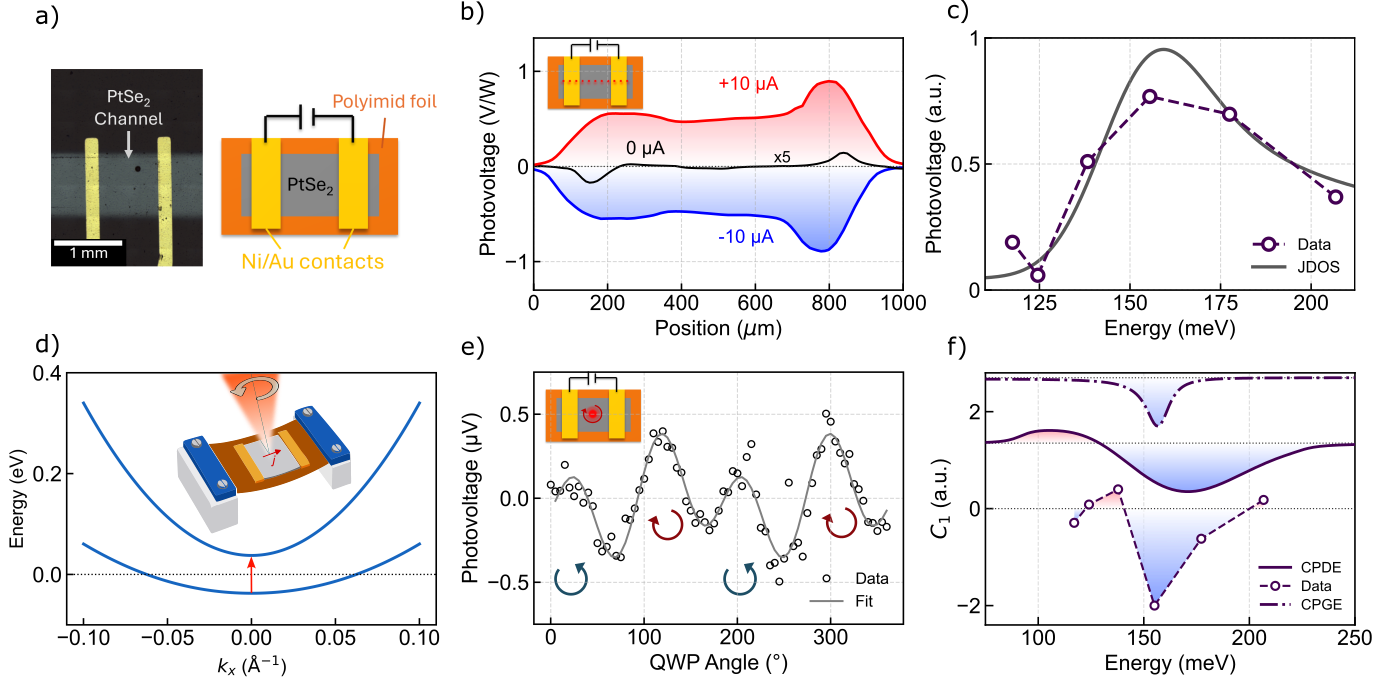}
	\caption{\textbf{Mid-IR photoresponse from the $K$-point gap} - a) Optical microscopy image and schematic illustration of the PtSe$_2$ channel on 25~$\upmu$m thick polyimide foil, contacted via Ni/Au contact pads. b) Photoresponse of the device along the channel for different current bias conditions of -10~$\upmu$A (red curve), 0~$\upmu$A (yellow curve, multiplied by a factor of 5 for better visibility) and +10~$\upmu$A (blue curve). c) Photoconductivity spectrum, recorded at +10~$\upmu$A bias current. The gray line depicts the spectral shape of the JDOS for optical transitions across the band gap at the $K$-point. d) Calculated band structure of a set of parabolic bands at the $K$-point, indicating possible optical transitions in the mid-IR. Inset: excitation geometry for polarization resolved measurements. e) Polarization resolved photoresponse, recorded in the center of the channel at off-normal incidence angle and without bias current, as a function of quarter-waveplate (QWP) angle and corresponding sinusoidal fit of $C_1 \sin 2\theta + L_1 \sin 4\theta + L_2 \cos 4\theta$. f) Experimental spectrum of the helicity dependent component $C_1$ as a function of laser energy and theoretical CPGE and CPDE spectrum arising from asymmetrical excitation of a toy-model band structure via circularly polarized light.}
    \label{fig:fig2}
\end{figure}

\noindent An optical microscopy image and schematic drawing of our device structure and geometry are provided in Fig.~\ref{fig:fig2}a.
For photoconductance experiments, the PtSe$_2$ film was optically excitated  with a tunable IR laser with wavelengths ranging from $\sim$~6~$\upmu$m up to $\sim$~11~$\upmu$m. 

Under an applied bias current, we observe a pronounced photovoltage response across the entire length of the PtSe$_2$ channel which peaks at the gold-contact interface (Fig.~\ref{fig:fig2}b).
Inside the channel, the photovoltage scales linearly with applied bias current (see in Fig. S2a). 
Correspondingly, in the absence of bias current, photovoltage is observed only at the contacts and remains negligible within the channel.
This indicates that, while contact interfaces strongly influence the optoelectronic response in PtSe$_2$ devices, the center of the PtSe$_2$ film reflects the intrinsic material response to optical excitation.
By measuring the spectrally dependent photoconductance under bias, we obtain a measure for the mid-IR absorption spectrum of our PtSe$_2$ shown in Fig.~\ref{fig:fig2}c.
Spectrally resolved photovoltage measurements in the 110–210~meV range reveal a peak near 160~meV, which is consistent with SOC induced band splitting at the $K$-point (Fig.~\ref{fig:fig2}d).
This absorption maximum agrees well with the joint density of states (JDOS) for optical transitions between the pair of conduction bands. 
We note that a high resolution spectrum reveals oscillations originating from thin-film interference in the polyimide substrate (see Fig. S2b).

After establishing the absorption of mid-IR radiation via photoconductance measurements, we turn towards studying non-linear photocurrent mechanisms.
To this end, we study the PtSe$_2$ film in the unbiased state in order to reveal intrinsic polarization driven photocurrents under oblique angle of incidence (Fig.~\ref{fig:fig2}d inset).

To probe nonlinear polarization sensitivity in detail, the photocurrent was recorded as a function of the quarter-wave plate (QWP) rotation angle (Fig.~\ref{fig:fig2}e). The response shows sinusoidal modulation with both two-fold (helicity-dependent) and four-fold (linear polarization) components. Notably, a helicity-dependent nonlinear component of the photocurrent emerges under circularly polarized illumination. Following standard procedure, we fit the photocurrent, $I(\theta)$, as a function of QWP angle, $\theta$:
\begin{equation}
    I(\theta) = D + C_1 \sin(2\theta) + L_1 \sin(4\theta) + L_2 \cos(4\theta)~.
    \label{IQWP}
\end{equation}
\noindent The helicity-dependent term $C_1$, corresponds to the two-fold contribution, which reverses sign with the handedness of circular polarization. This behavior indicates the presence of second-order nonlinear processes, such as the circular photogalvanic effect (CPGE) and the circular photon drag effect (CPDE), where the former is nominally forbidden in centrosymmetric materials. The coefficients $D$, $L_1$ and $L_2$ denote polarization independent and linear polarization sensitive contributions such as linear PGE (LPGE) and PDE (LPDE), and dichroic effects to the photocurrent. 

Next we focus on the spectral dependence of the observed phenomena.
Fig.~\ref{fig:fig2}f shows the extracted circular component $C_1$ for different excitation energies.
We find a maximum in the spectral response in good agreement with the previously identified resonance at the $K$-point.
Moreover, we observe a sign change from positive $C_1$ to negative $C_1$ when the excitation energy is tuned through this resonance.
Such a bipolar dependence of the helical photocurrent on photon energy has been predicted to appear in the CPDE for materials that lack particle-hole symmetry between the conduction and valence bands \cite{Nagaosa}.
At the $K$-point in PtSe$_2$, the mass difference between conduction and valence bands fulfills this asymmetry criterion \cite{Li2017}.
Thick TAC synthesized PtSe$_2$ has also previously demonstrated CPGE at visible wavelengths, and this was argued to originate from inversion breaking potentials developing between the substrate and the top of the PtSe$_2$ film \cite{Hemmat2022}. 

To determine whether the source of the observed $C_1(\hbar \omega)$ behavior is CPGE or CPDE, we develop a simple anisotropic 3D Bernevig-Hughes-Zhang (BHZ) Hamiltonian description of the band structure at the $K$-point, and used this to compute the appropriate NLO tensor elements according to Refs. \cite{Nagaosa,Sharifpour2026}.
The polycrystallinity of TAC grown PtSe$_2$ endows the material with isotropic rotational symmetry, making the point group symmetry of the surface effectively $C_{\infty v}$ rather than $C_{3v}$ \cite{Heiserer2025,Connelly2024}.
For this highly restrictive symmetry, and light incident along the $y$-$z$ plane, both the PGE and PDE generate a photocurrent vector $\mathbf{j}$ according to:
\begin{equation}
\begin{pmatrix}
j_x \\
j_y
\end{pmatrix}
=
\begin{pmatrix}
C_1 \sin (2\theta) + L_1 (\sin 4\theta) \\
D + L_2 \cos (4\theta) 
\end{pmatrix}~.
\label{eq:current_components}
\end{equation}
\noindent The dependence of $C_1$ and $L_1$ on the PGE and PDE tensors $\sigma^{ijk}_{\text{PG}}$ and $\sigma^{ijkl}_{\text{PD}}$, and the subsequent calculations based on the BHZ Hamiltonian are discussed in the Supplemental Material. We note that PtSe$_2$ is inherently centrosymmetric, so the PGE response is automatically zero. We can account for the substrate induced inversion breaking discussed in previous work by incorporating a Rashba gap at the $K$-point (see Supplementary Material in Fig. S3a), activating the photogalvanic response. The results for our toy-model Hamiltonian are shown in Fig.~\ref{fig:fig2}f. The CPDE response is seen to be bipolar like the experimental data, this is due to the anisotropy between inter and intralayer coefficients of the Hamiltonian. The CPGE response, on the other hand, is unipolar and is dictated by the strength of the Rashba gap. We show additionally in the Supplemental that for varying chemical potentials the CPDE response retains its form, but the relative strength of the two peaks varies (Fig. S4b). For varying chemical potential the location of the CPGE peak moves (Fig. S3b). Altogether, given the character of the calculated PGE and PDE responses, we conclude that our helical photocurrent is dominated by a PDE response. While PGE can always be present due to the surface or substrate inversion breaking fields, given that PDE is a bulk response and present in centrosymmetric materials, it is natural for it to dominate the NLO response here.

Note the form of Eq. \eqref{eq:current_components} would indicate that for our experimental set-up we should only expect to measure a $C_1$ and $L_1$ contribution to $I_x(\theta)$ in Fig.~\ref{fig:fig2}e. The presence of $D$ and $L_2$ are, however, unsurprising. An offset photothermalelectric current is almost always present in real device measurements unless care is taken to perfectly center the beam between contacts \cite{Kastl2015}. The presence of $L_2$ can be explained by the intermixing of $j_y$ and $j_x$ photocurrent densities due to electrical boundary conditions, and any linear (or s versus p) dichroism coupled to the photothermal electric response \cite{Levitov2014,Luo2022}.

\subsection*{Strain tuning of helicity dependent photocurrents}
\label{results:3}
\begin{figure}
	\centering
	\includegraphics[width=1\textwidth]{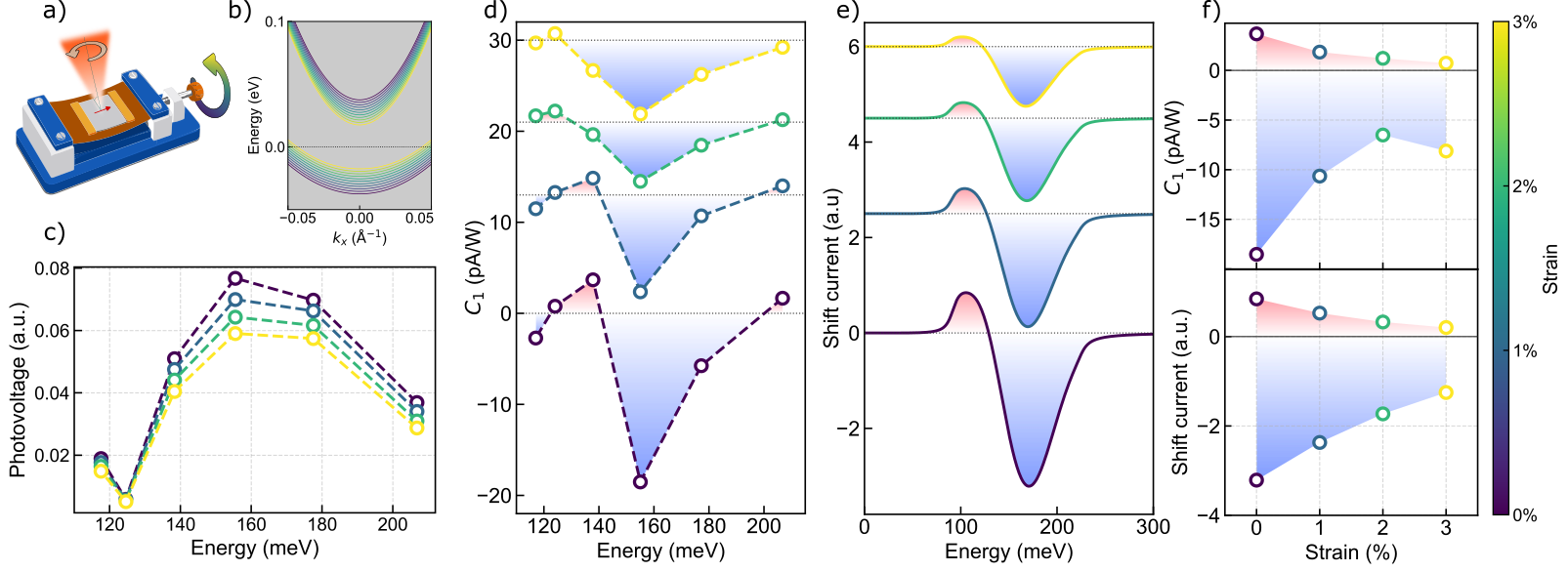}
	\caption{\textbf{Strain-tunable photoresponse} - The amount of strain is indicated by the global color scale on the far right of this figure. a) Schematic of the measurement geometry, where uniaxial tensile strain is applied in parallel direction to the electronic channel and orthogonal to the off-normal incidence of polarized light. b) Calculated deformation of the toy-model band structure around the $K$-point under uniaxial tensile strain. c) Photoconductivity spectrum under +10~$\upmu$A bias current for 0$\%$ to 3$\%$ tensile strain.
    d) Helicity dependent photocurrent $C_1$ as a function of photon energy and for increasing tensile strain. The curves for different strain values are offset for clarity. e) Calculated CPDE as a function of photon energy and for increasing tensile strain. The curves for different strain values are offset for clarity. f) Maximum positive and maximum negative helicity dependent photocurrents as a function of strain as extracted from experiment (top) and theory (bottom).}
    \label{fig:fig3}
\end{figure}

\noindent To investigate the impact of mechanical deformation on the photoresponse, we measured both photoconductivity and non-linear photocurrent spectroscopy while applying uniaxial strain to the PtSe$_2$ film parallel to the transport direction (Fig.~\ref{fig:fig3}a).
To first-order, strain will modify the band-gap at the $K$-point as can be seen in our toy-model (Fig.~\ref{fig:fig3}b). 
Experimentally we find that the overall photoresponse amplitude decreases with increasing tensile strain.
In the case of the photovoltage under bias, which we interpret to be proportional to the absorption driven change in channel resistance, we find that the spectral shape is mostly unaffected (Fig.~\ref{fig:fig3}c).
The monotonic amplitude change is consistent with the strain-induced overall change in PtSe$_2$ channel resistance, owing to its pronounced negative piezoresistivity.

Generally, local photocurrents are expected to scale with the channel resistance according to a formulation of the Shockley-Ramo theorem \cite{Levitov2014, Seifert2019}.
However, our observed helicity dependent component of the polarization resolved photocurrent depicts a  nmore pronounced change in amplitude, which cannot be explained by a piezoresistive channel resistance alone, and which also varies in spectral shape under deformation (Fig.~\ref{fig:fig3}d):
The overall amplitude decreases by more than 50$\%$ at 3$\%$ tensile strain, compared with the expected 21$\%$ contribution from a gauge factor of $\sim$~7, and the resonance peak broadens with increasing strain. (Fig.~\ref{fig:fig3}d)
At off-resonance energies, the coefficient $C_1$ even changes sign under strain (Fig.~\ref{fig:fig3}d).

We can augment the theoretical investigation of CPDE in the previous discussion to include strain deformation potentials that modify the Hamiltonian parameters \cite{Bir1974}. Exact determination of deformation potentials requires extensive density functional theory calculations, or strain based studies, so here we simply assume that to first-order strain will modify the band-gap at the $K$-point. In the Supplementary Material we show that by coupling the gap at the $K$-point, $M_0$, to total strain, $\epsilon= \epsilon_{xx} + \epsilon_{yy} + \epsilon_{zz}$, via a negative deformation potential, $\Lambda$, as $M(\epsilon) = M_0 + \Lambda \epsilon$, that the CPDE response weakens (see Fig.~\ref{fig:fig3}e), and the relative strength of the positive to the negative peak decreases with increasing tensile strain, in good agreement with our experimental observation (Fig.~\ref{fig:fig3}f). 
We mention that also our extracted coefficients $L_1$ and $L_2$ in principle match the theoretical expectations for a LPDE response as depicted in the Supplementary Material (see in Fig. S4a and S5). We note, however, that additional sources of linear polarization dependence could be present, such as photothermal electric effect with dichroic absorption, or interference and reflectance effects due to optical elements and off-normal incidence geometry. Thus, a linear polarization dependence cannot primarily be attributed to nonlinear photocurrent contributions. Overall, our findings reveal that strain provides a useful tuning parameter, controlling both the magnitude and direction of helicity-dependent photocurrents.

\section*{DISCUSSION}

\noindent We have demonstrated that the optoelectronic response of PtSe$_2$ thin films can be tuned by mechanical strain.
The films, synthesized on flexible substrates, exhibit piezoresistive behavior as well as a pronounced polarization sensitive nonlinear optical response.
Theory suggests that a circular photon drag effect is the dominant underlying mechanism for occurring helicity dependent photocurrents.
In good agreement with the theory, the amplitude and response spectrum can be tuned \textit{in-situ} via tensile strain, establishing strain as an effective tuning knob for optoelectronic nonlinearities in 2D semimetals.
Our work contributes to the fundamental understanding of nonlinear photocurrents in layered semimetals and reveals a route to integrate PtSe$_2$ into flexible, reconfigurable and polarization-sensitive optoelectronic platforms.
Potential applications include tunable photodetectors, polarization-sensitive imaging, and mechanically reconfigurable nonlinear optical devices.

\section*{MATERIALS AND METHODS}


\subsection*{Film synthesis and device fabrication}
Platinum-diselenide thin films were synthesized directly on 25 $\upmu m$ polyimide foils via a low-temperature thermally assisted conversion (TAC) process \cite{Boland2019}.
First, patterned platinum thin films were deposited by sputtering on polyimide substrates using a Cressington 108auto sputter coater.
The samples with the pre-deposited platinum films were subsequently selenized in a controlled selenium rich environment at 400~°C for 120~minutes, producing continuous polycrystalline PtSe$_2$ films with sputtering-time dependent thickness, in this case 10-12~nm.
To define electrical contacts for optoelectronic measurements, we used evaporated Ni/Au metal pads, patterned either via a shadow mask or standard photolithography.
This approach gives us reproducible films directly on a flexible substrate.

\subsection*{Experimental setup}

A dedicated optoelectronic mid-IR spectroscopy platform was developed that allows simultaneous spectroscopy and electrical transport measurements under tunable mechanical deformation of the sample.

The structured polyimide foil was installed into a dedicated vice with 4 point clamp.
This system allows manual control of uniaxial strain, applied directly to the flexible substrate close to the PtSe$_2$ channel.
The optoelectronic response was measured under illumination of tunable infrared lasers (DayLight Solutions MIRcat QT) with wavelength ranging from 5.7~$\upmu$m to 11~$\upmu$m.

The laser was modulated using an optical chopper wheel operating at approximately 70~Hz, providing a reference signal for the lock-in amplifier (Stanford SRS860). 
The linear polarization of the IR laser beam was initially defined by a linear polarizer and subsequently rotated using quarter-wave plates mounted on a Thorlabs motorized rotation stage to create circularly polarized light.
The combination of the linear polarizer and quarter-wave plates allowed precise control over both the polarization angle and the helicity of the incident beam.
The power of the laser was measured using a Thorlabs InGaAs Free-Space Amplified Photodetector.

Spatially resolved photocurrent mapping was performed using a piezo scanning stage enabling us to probe local variations in the optoelectronic response.
Data acquisition and hardware control were implemented using Python, including synchronization of linear and rotational stages, as well as signal readout.

Raman spectra were recorded using a WITec Alpha 300 confocal Raman microscope.
A 532~nm laser with a power of 0.5~mW was employed for all measurements in combination with a 100× objective lens.
A diffraction grating with 2400 grooves per mm was used for spectral dispersion.
Each spectrum presented represents the average of several acquisition points.
Post-processing and data analysis were carried out using Python.


\clearpage 

%
\bibliography{SciAdv_Temp/sample_test} 

@article{Hemmat2023,
author = {Hemmat, Minoosh and Ayari, Sabrine and Mičica, Martin and Vergnet, Hadrien and Guo, Shasha and Arfaoui, Mehdi and Yu, Xuechao and Vala, Daniel and Wright, Adrien and Postava, Kamil and Mangeney, Juliette and Carosella, Francesca and Jaziri, Sihem and Wang, Qi Jie and Liu, Zheng and Tignon, Jérôme and Ferreira, Robson and Baudin, Emmanuel and Dhillon, Sukhdeep},
title = {Layer-controlled nonlinear terahertz valleytronics in two-dimensional semimetal and semiconductor PtSe$_2$},
journal = {InfoMat},
volume = {5},
number = {11},
pages = {e12468},
doi = {https://doi.org/10.1002/inf2.12468},
url = {https://onlinelibrary.wiley.com/doi/abs/10.1002/inf2.12468},
eprint = {https://onlinelibrary.wiley.com/doi/pdf/10.1002/inf2.12468},
year = {2023}
}

@article{Wang2015,
	title = {Monolayer {PtSe}$_{\textrm{2}}$ , a {New} {Semiconducting} {Transition}-{Metal}-{Dichalcogenide}, {Epitaxially} {Grown} by {Direct} {Selenization} of {Pt}},
	volume = {15},
	issn = {1530-6984, 1530-6992},
	url = {https://pubs.acs.org/doi/10.1021/acs.nanolett.5b00964},
	doi = {10.1021/acs.nanolett.5b00964},
	language = {en},
	number = {6},
	urldate = {2025-11-25},
	journal = {Nano Lett.},
	author = {Wang, Yeliang and Li, Linfei and Yao, Wei and Song, Shiru and Sun, J. T. and Pan, Jinbo and Ren, Xiao and Li, Chen and Okunishi, Eiji and Wang, Yu-Qi and Wang, Eryin and Shao, Yan and Zhang, Y. Y. and Yang, Hai-tao and Schwier, Eike F. and Iwasawa, Hideaki and Shimada, Kenya and Taniguchi, Masaki and Cheng, Zhaohua and Zhou, Shuyun and Du, Shixuan and Pennycook, Stephen J. and Pantelides, Sokrates T. and Gao, Hong-Jun},
	month = jun,
	year = {2015},
	pages = {4013--4018},
}

@article{Boland2019,
	title = {{PtSe}$_{\textrm{2}}$ grown directly on polymer foil for use as a robust piezoresistive sensor},
	volume = {6},
	issn = {2053-1583},
	url = {https://iopscience.iop.org/article/10.1088/2053-1583/ab33a1},
	doi = {10.1088/2053-1583/ab33a1},
	number = {4},
	urldate = {2026-01-22},
	journal = {2D Mater.},
	author = {Boland, Conor S and Ó Coileáin, Cormac and Wagner, Stefan and McManus, John B and Cullen, Conor P and Lemme, Max C and Duesberg, Georg S and McEvoy, Niall},
	month = aug,
	year = {2019},
	pages = {045029},
}

@article{Han2019,
	title = {Horizontal-to-{Vertical} {Transition} of {2D} {Layer} {Orientation} in {Low}-{Temperature} {Chemical} {Vapor} {Deposition}-{Grown} {PtSe}$_{\textrm{2}}$ and {Its} {Influences} on {Electrical} {Properties} and {Device} {Applications}},
	volume = {11},
	copyright = {https://doi.org/10.15223/policy-029},
	issn = {1944-8244, 1944-8252},
	url = {https://pubs.acs.org/doi/10.1021/acsami.9b01078},
	doi = {10.1021/acsami.9b01078},
	language = {en},
	number = {14},
	urldate = {2026-01-22},
	journal = {ACS Appl. Mater. Interfaces},
	author = {Han, Sang Sub and Kim, Jong Hun and Noh, Chanwoo and Kim, Jung Han and Ji, Eunji and Kwon, Junyoung and Yu, Seung Min and Ko, Tae-Jun and Okogbue, Emmanuel and Oh, Kyu Hwan and Chung, Hee-Suk and Jung, YounJoon and Lee, Gwan-Hyoung and Jung, Yeonwoong},
	month = apr,
	year = {2019},
	pages = {13598--13607},
}

@article{Chen2024,
	title = {Defect-induced helicity dependent terahertz emission in {Dirac} semimetal {PtTe}$_{\textrm{2}}$ thin films},
	volume = {15},
	issn = {2041-1723},
	url = {https://www.nature.com/articles/s41467-024-46821-8},
	doi = {10.1038/s41467-024-46821-8},
	language = {en},
	number = {1},
	urldate = {2026-01-22},
	journal = {Nat. Commun.},
	author = {Chen, Zhongqiang and Qiu, Hongsong and Cheng, Xinjuan and Cui, Jizhe and Jin, Zuanming and Tian, Da and Zhang, Xu and Xu, Kankan and Liu, Ruxin and Niu, Wei and Zhou, Liqi and Qiu, Tianyu and Chen, Yequan and Zhang, Caihong and Xi, Xiaoxiang and Song, Fengqi and Yu, Rong and Zhai, Xuechao and Jin, Biaobing and Zhang, Rong and Wang, Xuefeng},
	month = mar,
	year = {2024},
	pages = {2605},
}

@article{lemme_nanoelectromechanical_2020,
	title = {Nanoelectromechanical {Sensors} {Based} on {Suspended} {2D} {Materials}},
	volume = {2020},
	copyright = {http://creativecommons.org/licenses/by/4.0/},
	issn = {2639-5274},
	url = {https://spj.science.org/doi/10.34133/2020/8748602},
	doi = {10.34133/2020/8748602},
	language = {en},
	urldate = {2026-01-26},
	journal = {Research},
	author = {Lemme, Max C. and Wagner, Stefan and Lee, Kangho and Fan, Xuge and Verbiest, Gerard J. and Wittmann, Sebastian and Lukas, Sebastian and Dolleman, Robin J. and Niklaus, Frank and Van Der Zant, Herre S. J. and Duesberg, Georg S. and Steeneken, Peter G.},
	month = jan,
	year = {2020},
	pages = {2020/8748602},
}

@article{OBrien2016,
	title = {Raman characterization of platinum diselenide thin films},
	volume = {3},
	issn = {2053-1583},
	url = {https://iopscience.iop.org/article/10.1088/2053-1583/3/2/021004},
	doi = {10.1088/2053-1583/3/2/021004},
	number = {2},
	urldate = {2026-01-26},
	journal = {2D Mater.},
	author = {O'Brien, Maria and McEvoy, Niall and Motta, Carlo and Zheng, Jian-Yao and Berner, Nina C and Kotakoski, Jani and Elibol, Kenan and Pennycook, Timothy J and Meyer, Jannik C and Yim, Chanyoung and Abid, Mohamed and Hallam, Toby and Donegan, John F and Sanvito, Stefano and Duesberg, Georg S},
	month = apr,
	year = {2016},
	pages = {021004},
}

@article{Lan2020,
	title = {{2D} materials beyond graphene toward {Si} integrated infrared optoelectronic devices},
	volume = {12},
	issn = {2040-3364, 2040-3372},
	url = {https://xlink.rsc.org/?DOI=D0NR02574G},
	doi = {10.1039/D0NR02574G},
	language = {en},
	number = {22},
	urldate = {2026-01-26},
	journal = {Nanoscale},
	author = {Lan, Changyong and Shi, Zhe and Cao, Rui and Li, Chun and Zhang, Han},
	year = {2020},
	pages = {11784--11807},
}

@article{Cheng2019,
	title = {Recent {Advances} in {Optoelectronic} {Devices} {Based} on {2D} {Materials} and {Their} {Heterostructures}},
	volume = {7},
	issn = {2195-1071, 2195-1071},
	url = {https://advanced.onlinelibrary.wiley.com/doi/10.1002/adom.201800441},
	doi = {10.1002/adom.201800441},
	language = {en},
	number = {1},
	urldate = {2026-01-26},
	journal = {Adv. Opt. Mat.},
	author = {Cheng, Jinbing and Wang, Chunlan and Zou, Xuming and Liao, Lei},
	month = jan,
	year = {2019},
	pages = {1800441},
}

@article{Ponraj2016,
	title = {Photonics and optoelectronics of two-dimensional materials beyond graphene},
	volume = {27},
	issn = {0957-4484, 1361-6528},
	url = {https://iopscience.iop.org/article/10.1088/0957-4484/27/46/462001},
	doi = {10.1088/0957-4484/27/46/462001},
	number = {46},
	urldate = {2026-01-26},
	journal = {Nanotechnology},
	author = {Ponraj, Joice Sophia and Xu, Zai-Quan and Dhanabalan, Sathish Chander and Mu, Haoran and Wang, Yusheng and Yuan, Jian and Li, Pengfei and Thakur, Siddharatha and Ashrafi, Mursal and Mccoubrey, Kenneth and Zhang, Yupeng and Li, Shaojuan and Zhang, Han and Bao, Qiaoliang},
	month = nov,
	year = {2016},
	pages = {462001},
}

@article{Bhimanapati2015,
	title = {Recent {Advances} in {Two}-{Dimensional} {Materials} beyond {Graphene}},
	volume = {9},
	issn = {1936-0851, 1936-086X},
	url = {https://pubs.acs.org/doi/10.1021/acsnano.5b05556},
	doi = {10.1021/acsnano.5b05556},
	language = {en},
	number = {12},
	urldate = {2026-01-26},
	journal = {ACS Nano},
	author = {Bhimanapati, Ganesh R. and Lin, Zhong and Meunier, Vincent and Jung, Yeonwoong and Cha, Judy and Das, Saptarshi and Xiao, Di and Son, Youngwoo and Strano, Michael S. and Cooper, Valentino R. and Liang, Liangbo and Louie, Steven G. and Ringe, Emilie and Zhou, Wu and Kim, Steve S. and Naik, Rajesh R. and Sumpter, Bobby G. and Terrones, Humberto and Xia, Fengnian and Wang, Yeliang and Zhu, Jun and Akinwande, Deji and Alem, Nasim and Schuller, Jon A. and Schaak, Raymond E. and Terrones, Mauricio and Robinson, Joshua A.},
	month = dec,
	year = {2015},
	pages = {11509--11539},
}

@article{Chen2020,
	title = {{2D} layered noble metal dichalcogenides ({Pt}, {Pd}, {Se}, {S}) for electronics and energy applications},
	volume = {7},
	issn = {25900498},
	url = {https://linkinghub.elsevier.com/retrieve/pii/S2590049820300230},
	doi = {10.1016/j.mtadv.2020.100076},
	language = {en},
	urldate = {2026-01-26},
	journal = {Materials Today Advances},
	author = {Chen, E. and Xu, W. and Chen, J. and Warner, J.H.},
	month = sep,
	year = {2020},
	pages = {100076},
}

@article{Lukas2025,
	title = {Piezoresistive {Platinum} {Diselenide} {Pressure} {Sensors} with {Reliable} {High} {Sensitivity} and {Their} {Integration} into {Complementary} {Metal}-{Oxide}-{Semiconductor} {Circuits}},
	volume = {19},
	copyright = {https://creativecommons.org/licenses/by-nc-nd/4.0/},
	issn = {1936-0851, 1936-086X},
	url = {https://pubs.acs.org/doi/10.1021/acsnano.4c15098},
	doi = {10.1021/acsnano.4c15098},
	language = {en},
	number = {7},
	urldate = {2026-01-26},
	journal = {ACS Nano},
	author = {Lukas, Sebastian and Rademacher, Nico and Cruces, Sofía and Gross, Michael and Desgué, Eva and Heiserer, Stefan and Dominik, Nikolas and Prechtl, Maximilian and Hartwig, Oliver and Ó Coileáin, Cormac and Stimpel-Lindner, Tanja and Legagneux, Pierre and Rantala, Arto and Saari, Juha-Matti and Soikkeli, Miika and Duesberg, Georg S. and Lemme, Max C.},
	month = feb,
	year = {2025},
	pages = {7026--7037},
}

@article{Yim2018,
	title = {Electrical devices from top-down structured platinum diselenide films},
	volume = {2},
	issn = {2397-7132},
	url = {https://www.nature.com/articles/s41699-018-0051-9},
	doi = {10.1038/s41699-018-0051-9},
	language = {en},
	number = {1},
	urldate = {2026-01-26},
	journal = {npj 2D Mater. Appl.},
	author = {Yim, Chanyoung and Passi, Vikram and Lemme, Max C. and Duesberg, Georg S. and Ó Coileáin, Cormac and Pallecchi, Emiliano and Fadil, Dalal and McEvoy, Niall},
	month = feb,
	year = {2018},
	pages = {5},
}

@article{Yim2016,
	title = {High-{Performance} {Hybrid} {Electronic} {Devices} from {Layered} {PtSe}$_{\textrm{2}}$ {Films} {Grown} at {Low} {Temperature}},
	volume = {10},
	issn = {1936-0851, 1936-086X},
	url = {https://pubs.acs.org/doi/10.1021/acsnano.6b04898},
	doi = {10.1021/acsnano.6b04898},
	language = {en},
	number = {10},
	urldate = {2026-01-26},
	journal = {ACS Nano},
	author = {Yim, Chanyoung and Lee, Kangho and McEvoy, Niall and O’Brien, Maria and Riazimehr, Sarah and Berner, Nina C. and Cullen, Conor P. and Kotakoski, Jani and Meyer, Jannik C. and Lemme, Max C. and Duesberg, Georg S.},
	month = oct,
	year = {2016},
	pages = {9550--9558},
}

@article{Wagner2018,
	title = {Highly {Sensitive} {Electromechanical} {Piezoresistive} {Pressure} {Sensors} {Based} on {Large}-{Area} {Layered} {PtSe}$_{\textrm{2}}$ {Films}},
	volume = {18},
	copyright = {http://pubs.acs.org/page/policy/authorchoice\_termsofuse.html},
	issn = {1530-6984, 1530-6992},
	url = {https://pubs.acs.org/doi/10.1021/acs.nanolett.8b00928},
	doi = {10.1021/acs.nanolett.8b00928},
	language = {en},
	number = {6},
	urldate = {2026-01-26},
	journal = {Nano Lett.},
	author = {Wagner, Stefan and Yim, Chanyoung and McEvoy, Niall and Kataria, Satender and Yokaribas, Volkan and Kuc, Agnieszka and Pindl, Stephan and Fritzen, Claus-Peter and Heine, Thomas and Duesberg, Georg S. and Lemme, Max C.},
	month = jun,
	year = {2018},
	pages = {3738--3745},
}

@article{Prechtl2021,
	title = {Hybrid {Devices} by {Selective} and {Conformal} {Deposition} of {PtSe}$_{\textrm{2}}$ at {Low} {Temperatures}},
	volume = {31},
	issn = {1616-301X, 1616-3028},
	url = {https://advanced.onlinelibrary.wiley.com/doi/10.1002/adfm.202103936},
	doi = {10.1002/adfm.202103936},
	language = {en},
	number = {46},
	urldate = {2026-01-26},
	journal = {Adv. Funct. Mat.},
	author = {Prechtl, Maximilian and Parhizkar, Shayan and Hartwig, Oliver and Lee, Kangho and Biba, Josef and Stimpel‐Lindner, Tanja and Gity, Farzan and Schels, Andreas and Bolten, Jens and Suckow, Stephan and Giesecke, Anna Lena and Lemme, Max C. and Duesberg, Georg S.},
	month = nov,
	year = {2021},
	pages = {2103936},
}

@article{Heiserer2025,
	title = {Impact of {Strain} in {Free}‐{Standing} {PtSe}$_{\textrm{2}}$ in {Scalable} {2D} {MEMS}},
	volume = {37},
	issn = {0935-9648, 1521-4095},
	url = {https://advanced.onlinelibrary.wiley.com/doi/10.1002/adma.202412564},
	doi = {10.1002/adma.202412564},
	language = {en},
	number = {43},
	urldate = {2026-01-26},
	journal = {Advanced Materials},
	author = {Heiserer, Stefan and Galfe, Natalie and Loibl, Michael and Wagner, Maximilian and Hartwig, Oliver and Schlosser, Simon and Boche, Silke and Thornley, William and Clark, Nick and Lee, Kangho and Stimpel‐Lindner, Tanja and Ó Coileáin, Cormac and Kiendl, Josef and Haigh, Sarah J. and de Coster, George J. and Duesberg, Georg S. and Seifert, Paul},
	month = oct,
	year = {2025},
	pages = {e12564},
}

@article{Nagaosa,
	title = {Photon-drag photovoltaic effects and quantum geometric nature},
	volume = {122},
	issn = {0027-8424, 1091-6490},
	url = {https://pnas.org/doi/10.1073/pnas.2424294122},
	doi = {10.1073/pnas.2424294122},
	number = {9},
	urldate = {2026-01-26},
	journal = {Proc. Natl. Acad. Sci. U.S.A.},
	author = {Xie, Ying-Ming and Nagaosa, Naoto},
	month = mar,
	year = {2025},
	pages = {e2424294122},
}

@article{Ganichev,
	title = {Spin photocurrents in quantum wells},
	volume = {15},
	issn = {0953-8984, 1361-648X},
	url = {https://iopscience.iop.org/article/10.1088/0953-8984/15/20/204},
	doi = {10.1088/0953-8984/15/20/204},
	number = {20},
	urldate = {2026-01-26},
	journal = {J. Phys.: Condens. Matter},
	author = {Ganichev, S D and Prettl, W},
	month = may,
	year = {2003},
	pages = {R935--R983},
}

@article{Prechtl2025,
	title = {Scalable {Metal}–{Organic} {Chemical} {Vapor} {Deposition} of {High} {Quality} {PtSe}$_{\textrm{2}}$},
	volume = {11},
	issn = {2199-160X, 2199-160X},
	url = {https://advanced.onlinelibrary.wiley.com/doi/10.1002/aelm.202400392},
	doi = {10.1002/aelm.202400392},
	language = {en},
	number = {3},
	urldate = {2026-06-18},
	journal = {Adv. Elect. Mat.},
	author = {Prechtl, Maximilian and Heiserer, Stefan and Busch, Marc and Hartwig, Oliver and Ó Coileáin, Cormac and Stimpel‐Lindner, Tanja and Zhussupbekov, Kuanysh and Lee, Kangho and Zhussupbekova, Ainur and Berman, Samuel and Shvets, Igor V. and Duesberg, Georg S.},
	month = mar,
	year = {2025},
	pages = {2400392},
}

@article{HilseMBE,
	title = {Growth of ultrathin {Pt} layers and selenization into {PtSe}$_{\textrm{2}}$ by molecular beam epitaxy},
	volume = {7},
	issn = {2053-1583},
	url = {https://iopscience.iop.org/article/10.1088/2053-1583/ab9f91},
	doi = {10.1088/2053-1583/ab9f91},
	number = {4},
	urldate = {2026-06-18},
	journal = {2D Mater.},
	author = {Hilse, Maria and Wang, Ke and Engel-Herbert, Roman},
	month = oct,
	year = {2020},
	pages = {045013},
}

@article{MakAndHeinz2010,
	title = {Atomically {Thin} {MoS}$_{\textrm{2}}$ : {A} {New} {Direct}-{Gap} {Semiconductor}},
	volume = {105},
	copyright = {http://link.aps.org/licenses/aps-default-license},
	issn = {0031-9007, 1079-7114},
	shorttitle = {Atomically {Thin} {MoS} 2},
	url = {https://link.aps.org/doi/10.1103/PhysRevLett.105.136805},
	doi = {10.1103/PhysRevLett.105.136805},
	language = {en},
	number = {13},
	urldate = {2026-06-22},
	journal = {Phys. Rev. Lett.},
	author = {Mak, Kin Fai and Lee, Changgu and Hone, James and Shan, Jie and Heinz, Tony F.},
	month = sep,
	year = {2010},
	pages = {136805},
}

@inproceedings{Hemmat2022,
	address = {Delft, Netherlands},
	title = {Ultrafast {Terahertz} {Photocurrents} in {Semi}-metal and semiconductor few monolayer {PtSe}$_{\textrm{2}}$},
	copyright = {https://doi.org/10.15223/policy-029},
	isbn = {978-1-7281-9427-1},
	url = {https://ieeexplore.ieee.org/document/9895953/},
	doi = {10.1109/IRMMW-THz50927.2022.9895953},
	urldate = {2026-06-22},
	booktitle = {2022 47th {International} {Conference} on {Infrared}, {Millimeter} and {Terahertz} {Waves} ({IRMMW}-{THz})},
	author = {Hemmat, M. and Vergnet, H. and Ferreira, R. and Mangeney, J. and Yu, X. and He, Y. and Liu, Z. and Wang, Q. and Tignon, J. and Baudin, E. and Dhillon, S.},
	month = aug,
	year = {2022},
	note = {Journal Abbreviation: IEEE},
	pages = {1--2},
}

@article{Chen2025,
	title = {Giant {Photogalvanic} {Effect}-{Induced} {Terahertz} {Wave} {Emission} in {Wafer}-{Scale} {Type}-{II} {Dirac} {Semimetal} {PtTe}$_{\textrm{2}}$},
	volume = {17},
	copyright = {https://doi.org/10.15223/policy-029},
	issn = {1944-8244, 1944-8252},
	url = {https://pubs.acs.org/doi/10.1021/acsami.4c17117},
	doi = {10.1021/acsami.4c17117},
	language = {en},
	number = {2},
	urldate = {2026-06-22},
	journal = {ACS Appl. Mater. Interfaces},
	author = {Chen, Mingyi and Jiang, Tianran and Wang, Jiali and He, Zhihao and Wu, Huiping and Li, Jibin and Guo, Xinhao and Li, Wanjiong and Liu, Pingwei and Chen, Huanjun and Yu, Peng and Chen, Xinman and Sou, Iam Keong and Lai, Tianshu and Li, Shuwei and Chen, Ke and Wu, Shuxiang},
	month = jan,
	year = {2025},
	pages = {4137--4146},
}

@article{Yadav2024,
	title = {Highly {Efficient} {Spintronic} {Terahertz} {Emitter} {Utilizing} a {Large} {Spin} {Hall} {Conductivity} of {Type}-{II} {Dirac} {Semimetal} {PtTe}$_{\textrm{2}}$},
	volume = {24},
	copyright = {https://doi.org/10.15223/policy-029},
	issn = {1530-6984, 1530-6992},
	url = {https://pubs.acs.org/doi/10.1021/acs.nanolett.3c04986},
	doi = {10.1021/acs.nanolett.3c04986},
	language = {en},
	number = {7},
	urldate = {2026-06-22},
	journal = {Nano Lett.},
	author = {Yadav, Pinki and Xinhou, Chen and Bhatt, Shubham and Das, Samaresh and Yang, Hyunsoo and Mishra, Rahul},
	month = feb,
	year = {2024},
	pages = {2376--2383},
}

@article{Guo2020,
	title = {Anisotropic ultrasensitive {PdTe}$_{\textrm{2}}$ -based phototransistor for room-temperature long-wavelength detection},
	volume = {6},
	copyright = {https://creativecommons.org/licenses/by-nc/4.0/},
	issn = {2375-2548},
	url = {https://www.science.org/doi/10.1126/sciadv.abb6500},
	doi = {10.1126/sciadv.abb6500},
	language = {en},
	number = {36},
	urldate = {2026-06-22},
	journal = {Sci. Adv.},
	author = {Guo, Cheng and Hu, Yibin and Chen, Gang and Wei, Dacheng and Zhang, Libo and Chen, Zhiqingzi and Guo, Wanlong and Xu, Huang and Kuo, Chia-Nung and Lue, Chin Shan and Bo, Xiangyan and Wan, Xiangang and Wang, Lin and Politano, Antonio and Chen, Xiaoshuang and Lu, Wei},
	month = sep,
	year = {2020},
	pages = {eabb6500},
}

@article{DeCoster2026,
	title = {Visible and terahertz nonlinear responses in the topological noble metal dichalcogenide {PdTe}$_{\textrm{2}}$},
	volume = {113},
	issn = {2469-9950, 2469-9969},
	url = {https://link.aps.org/doi/10.1103/x5kl-blc5},
	doi = {10.1103/x5kl-blc5},
	language = {en},
	number = {16},
	urldate = {2026-06-22},
	journal = {Phys. Rev. B},
	author = {de Coster, George J. and Lafeta, Lucas and Heiserer, Stefan and Ó Coileáin, Cormac and Sofer, Zdenek and Hartschuh, Achim and Duesberg, Georg S. and Seifert, Paul},
	month = apr,
	year = {2026},
	pages = {165419},
}

@article{Liu2025,
	title = {Strain engineering of photoinduced anomalous {Hall} effect in topological insulator {Sb$_{\textrm{2}}$Te$_{\textrm{3}}$}},
	volume = {127},
	issn = {0003-6951, 1077-3118},
	url = {https://pubs.aip.org/apl/article/127/5/053102/3357735/Strain-engineering-of-photoinduced-anomalous-Hall},
	doi = {10.1063/5.0279026},
	language = {en},
	number = {5},
	urldate = {2026-06-22},
	journal = {Applied Physics Letters},
	author = {Liu, Tengfei and Hong, Xiyu and Lin, Zongkai and Qiu, Jiayi and Cheng, Shuying and Lai, Yunfeng and Chen, Yonghai and He, Ke and Yu, Jinling},
	month = aug,
	year = {2025},
	pages = {053102},
}

@article{Levitov2014,
	title = {Shockley-{Ramo} theorem and long-range photocurrent response in gapless materials},
	volume = {90},
	copyright = {http://link.aps.org/licenses/aps-default-license},
	issn = {1098-0121, 1550-235X},
	url = {https://link.aps.org/doi/10.1103/PhysRevB.90.075415},
	doi = {10.1103/PhysRevB.90.075415},
	language = {en},
	number = {7},
	urldate = {2026-06-22},
	journal = {Phys. Rev. B},
	author = {Song, Justin C. W. and Levitov, Leonid S.},
	month = aug,
	year = {2014},
	pages = {075415},
}

@article{Seifert2019,
	title = {Quantized {Conductance} in {Topological} {Insulators} {Revealed} by the {Shockley}-{Ramo} {Theorem}},
	volume = {122},
	issn = {0031-9007, 1079-7114},
	url = {https://link.aps.org/doi/10.1103/PhysRevLett.122.146804},
	doi = {10.1103/PhysRevLett.122.146804},
	language = {en},
	number = {14},
	urldate = {2026-06-22},
	journal = {Phys. Rev. Lett.},
	author = {Seifert, Paul and Kundinger, Marinus and Shi, Gang and He, Xiaoyue and Wu, Kehui and Li, Yongqing and Holleitner, Alexander and Kastl, Christoph},
	month = apr,
	year = {2019},
	pages = {146804},
}

@article{Cheng2023,
	title = {Giant photon momentum locked {THz} emission in a centrosymmetric {Dirac} semimetal},
	volume = {9},
	issn = {2375-2548},
	url = {https://www.science.org/doi/10.1126/sciadv.add7856},
	doi = {10.1126/sciadv.add7856},
	language = {en},
	number = {1},
	urldate = {2026-06-22},
	journal = {Sci. Adv.},
	author = {Cheng, Liang and Xiong, Ying and Kang, Lixing and Gao, Yu and Chang, Qing and Chen, Mengji and Qi, Jingbo and Yang, Hyunsoo and Liu, Zheng and Song, Justin C.W. and Chia, Elbert E. M.},
	month = jan,
	year = {2023},
	pages = {eadd7856},
}

@article{Parker2019,
	title = {Diagrammatic approach to nonlinear optical response with application to {Weyl} semimetals},
	volume = {99},
	url = {https://link.aps.org/doi/10.1103/PhysRevB.99.045121},
	doi = {10.1103/PhysRevB.99.045121},
	number = {4},
	journal = {Phys. Rev. B},
	publisher = {American Physical Society},
	author = {Parker, Daniel E. and Morimoto, Takahiro and Orenstein, Joseph and Moore, Joel E.},
	month = jan,
	year = {2019},
	pages = {045121},
}

@book{Bir1974,
	title = {Symmetry and strain-induced effects in semiconductors},
	publisher = {Wiley},
	author = {Bir, Gennadi{\u \i} Levikovich and Pikus, Grigori{\u \i} Ezekielevich},
	year = {1974},
}

@article{Li2017,
	title = {Topological origin of the type-{II} {Dirac} fermions in {PtSe}$_{\textrm{2}}$},
	volume = {1},
	url = {https://link.aps.org/doi/10.1103/PhysRevMaterials.1.074202},
	doi = {10.1103/PhysRevMaterials.1.074202},
	number = {7},
	journal = {Phys. Rev. Mater.},
	publisher = {American Physical Society},
	author = {Li, Yiwei and Xia, Yunyouyou and Ekahana, Sandy Adhitia and Kumar, Nitesh and Jiang, Juan and Yang, Lexian and Chen, Cheng and Liu, Chaoxing and Yan, Binghai and Felser, Claudia and Li, Gang and Liu, Zhongkai and Chen, Yulin},
	month = dec,
	year = {2017},
	pages = {074202},
}

@article{sharifpour2026,
  title = {Enhanced detection of circularly polarized photons with topological materials},
  author = {Sharifpour, Hamideh and Ghosh, Avik W. and de Coster, George J.},
  journal = {Phys. Rev. B},
  volume = {114},
  issue = {6},
  pages = {065308},
  numpages = {19},
  year = {2026},
  month = {Jul},
  publisher = {American Physical Society},
  doi = {10.1103/wy64-ngn2},
  url = {https://link.aps.org/doi/10.1103/wy64-ngn2}
}

@article{Connelly2024,
	title = {Emergence of threefold symmetric helical photocurrents in epitaxial low twinned {Bi}$_{\textrm{2}}$ {Se}$_{\textrm{3}}$},
	volume = {121},
	issn = {1091-6490},
	url = {http://dx.doi.org/10.1073/pnas.2307425121},
	doi = {10.1073/pnas.2307425121},
	number = {5},
	journal = {Proceedings of the National Academy of Sciences},
	publisher = {Proceedings of the National Academy of Sciences},
	author = {Connelly, Blair C. and Taylor, Patrick J. and de Coster, George J.},
	month = jan,
	year = {2024},
}

@article{Kastl2015,
	title = {Ultrafast helicity control of surface currents in topological insulators with near-unity fidelity},
	volume = {6},
	issn = {2041-1723},
	url = {http://dx.doi.org/10.1038/ncomms7617},
	doi = {10.1038/ncomms7617},
	number = {1},
	journal = {Nat. Commun.},
	publisher = {Springer Science and Business Media LLC},
	author = {Kastl, Christoph and Karnetzky, Christoph and Karl, Helmut and Holleitner, Alexander W.},
	month = mar,
	year = {2015},
}

@article{Luo2022,
	title = {Network analysis of {Weyl} semimetal photogalvanic systems},
	volume = {47},
	url = {https://opg.optica.org/ol/abstract.cfm?URI=ol-47-10-2450},
	doi = {10.1364/OL.452929},
	number = {10},
	journal = {Opt. Lett.},
	publisher = {Optica Publishing Group},
	author = {Luo, Haokun and Jia, Yufei and Tian, Fugu and Khajavikhan, Mercedeh and Christodoulides, Demetrios},
	month = may,
	year = {2022},
	pages = {2450--2453},
}

@article{Pierre2023,
	title = {Probing {Carrier} {Dynamics} in {Large}-{Scale} {MBE}-{Grown} {PtSe2} {Films} by {Terahertz} {Spectroscopy}},
	volume = {15},
	url = {https://doi.org/10.1021/acsami.3c09792},
	doi = {10.1021/acsami.3c09792},
	number = {44},
	journal = {ACS Appl. Mater. Interfaces},
	author = {Ji, Jie and Zhou, Yingqiu and Zhou, Binbin and Desgué, Eva and Legagneux, Pierre and Jepsen, Peter Uhd and Bøggild, Peter},
	year = {2023},
	pages = {51319--51329},
}

@article{Ansari2019,
	title = {Quantum confinement-induced semimetal-to-semiconductor evolution in large-area ultra-thin {PtSe}$_{\textrm{2}}$ films grown at 400 °{C}},
	volume = {3},
	issn = {2397-7132},
	url = {https://doi.org/10.1038/s41699-019-0116-4},
	doi = {10.1038/s41699-019-0116-4},
	number = {1},
	journal = {npj 2D Mater. Appl.},
	author = {Ansari, Lida and Monaghan, Scott and McEvoy, Niall and Ó Coileáin, Cormac and Cullen, Conor P. and Lin, Jun and Siris, Rita and Stimpel-Lindner, Tanja and Burke, Kevin F. and Mirabelli, Gioele and Duffy, Ray and Caruso, Enrico and Nagle, Roger E. and Duesberg, Georg S. and Hurley, Paul K. and Gity, Farzan},
	month = sep,
	year = {2019},
	pages = {33},
}

@article{Lee2013,
	title = {High-{Performance} {Sensors} {Based} on {Molybdenum} {Disulfide} {Thin} {Films}},
	volume = {25},
	url = {https://advanced.onlinelibrary.wiley.com/doi/abs/10.1002/adma.201303230},
	doi = {https://doi.org/10.1002/adma.201303230},
	number = {46},
	journal = {Adv. Mat.},
	author = {Lee, Kangho and Gatensby, Riley and McEvoy, Niall and Hallam, Toby and Duesberg, Georg S.},
	year = {2013},
	pages = {6699--6702},
}

@book{boyd2008nonlinear,
  author = {Boyd, Robert W.},
  title = {Nonlinear Optics (Third Edition)},
  publisher = {Academic Press},
  address = {Burlington},
  year = {2008},
  doi = {10.1016/B978-0-12-369470-6.00001-0}
}

\bibliographystyle{sciencemag}

%
%
%
%
%
%


\section*{Acknowledgments}
The authors thank C. Ó Coileáin for insightful discussions, careful reading of the manuscript, valuable comments, and assistance with language editing. 

\paragraph*{Funding:}
This work was supported by dtec.bw—Digitalization and Technology Research Center of the Bundeswehr through the project VITAL-SENSE.
dtec.bw is funded via the German Recovery and Resilience Plan by the European Union (NextGenerationEU).
This research was supported by the Army Research Office and was accomplished under Cooperative Agreement No. W911NF2520010 (STEP-TWO). The views and conclusions contained in this document are those of the authors and should not be interpreted as representing the official policies, either expressed or implied, of the Army Research Office or the U.S. Government. The U.S. Government is authorized to reproduce and distribute reprints for Government purposes notwithstanding any copyright notation thereon. 
\paragraph*{Author contributions:}
M.~G. and P.~S. conceived the research project and designed the experiments. M.~G. performed the IR photocurrent spectroscopy experiments. G.~J.~d.~C. carried out the theoretical calculations and developed the theoretical framework. J.~P. performed the Raman spectroscopy measurements. S.~H. synthesized the material. S.~S assisted with the fabrication of the strain device.  M.~G., G.~J.~d.~C. and P.~S. analyzed the data and interpreted the results. M.~G., P.~S. and G.~S.~D. wrote the manuscript with contributions from all authors.
\paragraph*{Competing interests:}
The authors declare no competing interests.
\paragraph*{Data, code and materials availability:}
All data needed to evaluate the conclusions in the paper are presented in the paper and/or the Supplementary Materials. The data is available from the authors upon reasonable request.

\subsection*{Supplementary materials}
Supplementary Text\\
Figs. S1 to S5\\
References \textit{33,43}\\ 


\newpage


\renewcommand{\thefigure}{S\arabic{figure}}
\renewcommand{\thetable}{S\arabic{table}}
\renewcommand{\theequation}{S\arabic{equation}}
\renewcommand{\thepage}{S\arabic{page}}
\setcounter{figure}{0}
\setcounter{table}{0}
\setcounter{equation}{0}
\setcounter{page}{1} 


\begin{center}
\section*{Supplementary Materials for\\ \scititle}

M. Gerlei$^{1\ast}$,
G. J. de Coster$^{2,3}$,
J. Papp$^{1}$,
S. Heiserer$^{1}$,\and
S. Schlosser$^{1}$,
G. S. Duesberg$^{1}$,
P. Seifert$^{1\ast}$\\
\small$^\ast$Corresponding author. Email: martin.gerlei@unibw.de, paul.seifert@unibw.de\\
\end{center}

\subsubsection*{This PDF file includes:}
Supplementary Text\\
Figures S1 to S5\\

\newpage


\section*{Supplementary Text}
\subsection*{Calculating the Shift PDE Tensor}

The shift PDE current that can contribute to CPDE as derived by Xie and Nagaosa is \cite{Nagaosa}:
\begin{equation}
\sigma_{shift}^{\lambda i j k} = \frac{\pi e^3}{2\hbar^2 \omega} \sum_{m\neq n} f_{nm} W^k_{nm} \left[ (R^{i;\lambda}_{mn} - R^{i;j}_{nm})G^{\lambda j}_{mn} + i \partial_{k_i} G^{\lambda j}_{mn} \right] \delta(\omega_{nm} - \omega) + (j \leftrightarrow k)^* ~,
\label{shifteqn}
\end{equation}
\noindent Here, $W^k_{nm} = v^k_{nn} + v^{k}_{mm}$, $R^{i;j}_{nm} = r^i_{nn} - r^i_{mm} + i \partial_{k_i} \log (r^j_{nm})$, and the quantum metric tensor is $G^{\lambda j}_{mn} = r^\lambda_{mn} r^j_{nm}$. Note that $f_{nm}$ is the difference of Fermi-Dirac distributions $f_D(\epsilon_n) - f_D(\epsilon_m)$, and $\hbar \omega_{nm} = \epsilon_n - \epsilon_m$, where $\epsilon_n$ are the eigenvalues of the associated Hamiltonian.  The velocity matrix $v^k_{nm}$ is evaluated for the $k$-th direction and eigenvalues $n$ and $m$:
\begin{equation}
    v^k_{nm}(\bm{k}) = \langle m \bm{k} | \frac{\partial H(\bm{k})}{\hbar \partial k_k} | n \bm{k} \rangle~.
\end{equation}
\noindent The quantity $r^i_{nm}$ is the interband representation of the position operator, determined through the velocity matrix as:
\begin{equation}
    r^i_{nm}(\bm{k}) = \frac{\hbar v^i_{nm}(\bm{k})}{i (\epsilon_n(\bm{k}) -  \epsilon_m(\bm{k}))} = \frac{v^i_{nm}(\bm{k})}{i\omega_{nm}}~.
\end{equation}
\noindent The Dirac delta function is implemented numerically via a Lorentzian broadening:
\begin{equation}
    \delta(\omega_{nm} - \omega) = \frac{1}{\pi} \text{Im} \left[ \frac{1}{\omega_{nm}-\omega - i \eta} \right]~,
\end{equation}
\noindent where $\hbar \eta$ sets the self-energy/lifetime (typically set to $\hbar \eta = 1$ meV). We observe that the tensor bracket in Eq.~\eqref{shifteqn}, defined as $\mathcal{B}$, can be significantly simplified:
\begin{equation}
\mathcal{B} \equiv (R^{i;\lambda}_{mn} - R^{i;j}_{nm})G^{\lambda j}_{mn} + i \partial_{k_i} G^{\lambda j}_{mn}~.
\label{Beqn}
\end{equation}
\noindent Applying the product rule to $G^{\lambda j}_{mn}$ we obtain:
\begin{equation}
i \partial_{k_i} G^{\lambda j}_{mn} = i \partial_{k_i} (r^\lambda_{mn} r^j_{nm}) = i (\partial_{k_i} r^\lambda_{mn}) r^j_{nm} + i r^\lambda_{mn} (\partial_{k_i} r^j_{nm})~.\notag
\end{equation}
\noindent Then, explicitly computing the $RG$ terms in Eq. \eqref{Beqn}, and noting that $i \partial_{k_i} \log (r^j_{nm}) = i \frac{\partial_{k_i} r^j_{nm}}{r^j_{nm}}$, yields:
\begin{align}
R^{i;\lambda}_{mn} G^{\lambda j}_{mn} &= \left[ (r^i_{mm} - r^i_{nn}) + i \frac{\partial_{k_i} r^\lambda_{mn}}{r^\lambda_{mn}} \right] r^\lambda_{mn} r^j_{nm} = (r^i_{mm} - r^i_{nn}) r^\lambda_{mn} r^j_{nm} + i (\partial_{k_i} r^\lambda_{mn}) r^j_{nm}~, \notag\\
R^{i;j}_{nm} G^{\lambda j}_{mn} &= \left[ (r^i_{nn} - r^i_{mm}) + i \frac{\partial_{k_i} r^j_{nm}}{r^j_{nm}} \right] r^\lambda_{mn} r^j_{nm} = (r^i_{nn} - r^i_{mm}) r^\lambda_{mn} r^j_{nm} + i r^\lambda_{mn} (\partial_{k_i} r^j_{nm})~.\notag
\end{align}
\noindent Substituting the expanded components back into $\mathcal{B}$ we obtain:
\begin{align}
\mathcal{B} &= \left[ (r^i_{mm} - r^i_{nn}) r^\lambda_{mn} r^j_{nm} + i (\partial_{k_i} r^\lambda_{mn}) r^j_{nm} \right]  - \left[ (r^i_{nn} - r^i_{mm}) r^\lambda_{mn} r^j_{nm} + i r^\lambda_{mn} (\partial_{k_i} r^j_{nm}) \right] \nonumber \\
&\quad + \left[ i (\partial_{k_i} r^\lambda_{mn}) r^j_{nm} + i r^\lambda_{mn} (\partial_{k_i} r^j_{nm}) \right]~.\notag\\
&=i \left[ \partial_{k_i} r^\lambda_{mn} - i(r^i_{mm} - r^i_{nn}) r^\lambda_{mn} \right] r^j_{nm} \equiv 2 i r^\lambda_{mn;i} r^j_{nm}~,
\label{Bsimp}
\end{align}
\noindent where we have used the fact that the term in square brackets is exactly the generalized covariant derivative, $r^\lambda_{mn;i}$. Substituting the simplified $\mathcal{B}$ bracket back into Eq.~\eqref{shifteqn}, the final expression for the shift PDE tensor is:
\begin{equation}
\sigma_{shift}^{\lambda i j k} = \frac{\pi e^3}{\hbar^2 \omega} \sum_{m\neq n} f_{nm} W^k_{nm} \left[ i r^\lambda_{mn;i} r^j_{nm} \right] \delta(\omega_{nm} - \omega) + (j \leftrightarrow k)^*
\end{equation}
\noindent where $r^\lambda_{mn;i}$ is evaluated using the gauge-invariant two-band approximation, and $r^j_{nm} = v^j_{nm} / i\omega_{nm}$. This formulation is entirely free of phase derivatives and is highly stable for numerical lattice integration.

\subsection*{The Two-Band Approximation and Strict Gauge Invariance}
The final expression in Eq. \eqref{Bsimp} contains a momentum derivative of the position operator, which can obtain a spurious gauge dependence when evaluated numerically. To achieve a strictly gauge-invariant, computable form, we map the covariant derivative entirely to the velocity and inverse-mass operators using the sum rules derived from the Aversa-Sipe commutation relations:
\begin{equation}
r^\lambda_{mn;i} = \frac{w^{\lambda i}_{mn}}{i\omega_{mn}} - \frac{v^i_{mn}\Delta^\lambda_{mn} + v^\lambda_{mn}\Delta^i_{mn}}{i\omega_{mn}^2} + \frac{1}{\omega_{mn}} \sum_{l \neq m,n} \left( r^i_{ml} v^\lambda_{ln} - v^\lambda_{ml} r^i_{ln} \right)~,
\end{equation}
\noindent where $w^{\lambda i}_{mn} = \langle m | \partial_{k_i} \partial_{k_\lambda} H | n \rangle / \hbar^2$ is the second derivative matrix, and $\Delta^i_{mn} = v^i_{mm} - v^i_{nn}$. The final sum over $l\neq m,n$ can be neglected in a two band system. The BHZ Hamiltonian we will use to describe the $K$-point of PtSe$_2$ is a minimal $4 \times 4$ Hamiltonian (e.g., the spinful 3D BHZ model), and here optical transitions predominantly occur strictly between a single valence and conduction band pair.  Under this two-band assumption, the covariant derivative reduces to a purely analytic, gauge-invariant matrix evaluation:
\begin{equation}
r^\lambda_{mn;i} \approx \frac{w^{\lambda i}_{mn}}{i\omega_{mn}} - \frac{v^i_{mn}(v^\lambda_{mm} - v^\lambda_{nn}) + v^\lambda_{mn}(v^i_{mm} - v^i_{nn})}{i\omega_{mn}^2}
\end{equation}

\subsection*{Calculating The Injection PDE Tensor}

The photon-drag injection current derived by Xie and Nagaosa can be written, after neglecting explicit three-band resonant processes, as
\begin{align}
\sigma^{\lambda i j k}_{\rm inj}(\omega)
\simeq
\frac{\pi e^3}{\hbar^2}
\sum_{m\neq n}
f_{nm}
\Bigg[
\frac{1}{2\omega}\Delta^i_{nm}
\left(
W^k_{nm}G^{\lambda j}_{mn}
+
W^j_{nm}G^{\lambda k}_{mn}
\right)
-
\frac{1}{2}\Delta^i_{nm}
\partial_{k_\lambda}G^{kj}_{mn}
\Bigg]
\delta(\omega_{mn}-\omega)~.
\label{eq:XN_injection_start}
\end{align}
\noindent By using the definitions above we can write this as
\begin{eqnarray}
\sigma^{\lambda i j k}_{\rm inj}(\omega)
&=&
\frac{\pi e^3}{\hbar^2}
\sum_{m\neq n}
f_{nm}
\Delta^i_{nm}\delta(\omega_{mn}-\omega)\times\notag\\
&~& \Bigg[
\frac{1}{2\omega}
\left(
W^k_{nm}r^\lambda_{mn}r^j_{nm}
+
W^j_{nm}r^\lambda_{mn}r^k_{nm}
\right)
-
\frac{1}{2}
\left(
r^k_{mn;\lambda}r^j_{nm}
+
r^k_{mn}r^j_{nm;\lambda}
\right)
\Bigg]
~,
\label{eq:injection_covariant_r}
\end{eqnarray}
which is straightforward to implement numerically.

\subsection*{The 3D BHZ Hamiltonian}
At the $K$ and $K'$ points in the Brillouin zone for PtSe$_2$ we adopt a simple spinful 3D BHZ Hamiltonian to describe the bands:
\begin{equation}
    H_{K}(\mathbf{k}) = \begin{pmatrix}
    \epsilon(\mathbf{k}) + d_z(\mathbf{k}) & A_x k_x - i A_y k_y & 0 & A_z k_z \\
    A_x k_x + i A_y k_y & \epsilon(\mathbf{k})-d_z (\mathbf{k}) & A_z k_z & 0 \\
    0 & A_z k_z & \epsilon(\mathbf{k}) + d_z(\mathbf{k}) & -(A_x k_x + i A_y k_y ) \\
    A_z k_z & 0 & -(A_x k_x -i A_y k_y ) & \epsilon(\mathbf{k}) -d_z(\mathbf{k}) 
    \end{pmatrix}~,
\end{equation}
\noindent where the diagonal kinetic and mass terms are:
\begin{eqnarray}
\epsilon(\mathbf{k}) &=& C - D_x k_x^2 - D_y k_y^2 - D_z k_z^2~, \notag \\
d_z(\mathbf{k}) &=& M - B_x k_x^2 - B_y k_y^2 - B_z k_z^2~. \notag
\end{eqnarray}
\noindent The continuum $4\times4$ Hamiltonian can be equivalently parameterized using the Dirac matrices $\Gamma_0, \Gamma_1, \Gamma_2, \Gamma_3$:
\begin{equation}
H_{K}(\mathbf{k}) = \epsilon(\mathbf{k})I_{4} + d_z(\mathbf{k})\Gamma_0 + A_x k_x\Gamma_1 + A_y k_y\Gamma_2 + A_z k_z\Gamma_3
\end{equation}
\noindent where the matrices are explicitly defined using the Kronecker tensor product ($\otimes$) of the spin Pauli matrices ($\sigma_0, \sigma_x, \sigma_y, \sigma_z$) and the orbital/band Pauli matrices ($\tau_0, \tau_x, \tau_y, \tau_z$):
\begin{equation}
\Gamma_0 = \sigma_0 \otimes \tau_z ~, \quad \Gamma_1 = \sigma_z \otimes \tau_x ~, \quad \Gamma_2 = \sigma_0 \otimes \tau_y ~, \quad \Gamma_3 = \sigma_x \otimes \tau_x~.
\end{equation}
\noindent For the numerical simulations of the PtSe$_2$ PDE, the baseline parameters defining the unstrained Hamiltonian are set to: $A_x = A_y = -0.25$ eV$\cdot$\AA, $A_z = -1.25$ eV$\cdot$\AA, $B_x = B_y = -10.0$ eV$\cdot$\AA$^2$, $B_z = -12.0$ eV$\cdot$\AA$^2$, $D_x = D_y = D_z = -20.0$ eV$\cdot$\AA$^2$, $C = 0.0$ eV, and the initial mass gap $M_0 = 37.5$ meV. This produces the simple two-band band structure in Fig. 2d that phenomenologically captures the mass anisotropy of published PtSe$_2$ band structures. We note that at $K'$, which is the time reversal pair of $K$, the the Hamiltonian is:
\begin{equation}
    H_{K'}(k_x,k_y,k_z) = \Theta H_K(-k_x,-k_y,-k_z) \Theta^{-1} = H_K(k_x,-k_y,k_z)~,
\end{equation}
\noindent where we have used the standard $\Theta = i \sigma_y K$ time reversal operator.  All subsequent calculations are performed for both the $K$ and $K'$ and added together, and we generally find the photocurrent tensor at $K$ and $K'$ are the same.

\subsection*{Analytical Matrix Operators}
The momentum-space derivatives of the Hamiltonian define the velocity operator ($v$) and the inverse effective mass operator ($w$). For analytical evaluation, the first and second derivatives of the Hamiltonian evaluate exactly to:
\begin{eqnarray}
\hbar v^i(\mathbf{k}) = \partial_{k_i} H &=& -2D_i k_i I_{4} - 2B_i k_i \Gamma_0 + A_i \Gamma_i \\
\hbar^2 w^{ij}(\mathbf{k}) = \partial_{k_i} \partial_{k_j} H &=& -2 \delta_{ij} (D_i I_{4} + B_i \Gamma_0)
\end{eqnarray}
\noindent The energy eigenvalues for the conduction ($c$) and valence ($v$) bands are doubly degenerate:
\begin{eqnarray}
\epsilon_{c,v}(\mathbf{k}) &=& \epsilon(\mathbf{k}) \pm d(\mathbf{k})~,\notag \\
d(\mathbf{k}) &=& \sqrt{d_z^2(\mathbf{k}) + A_x^2 k_x^2 + A_y^2 k_y^2 + A_z^2 k_z^2}\notag
\end{eqnarray}
\noindent The terms $W^k_{cv}(\mathbf{k})$ and interband position matrix elements $r^j_{cv}(\mathbf{k}) $ ($c \neq v$) simplify to:
\begin{eqnarray}
W^k_{cv}(\mathbf{k}) &=& v^k_{cc}(\mathbf{k}) + v^k_{vv}(\mathbf{k}) = \frac{1}{\hbar} \partial_{k_k} (\epsilon_c + \epsilon_v) = \frac{1}{\hbar} \partial_{k_k} (2\epsilon(\mathbf{k})) = -\frac{4D_k k_k}{\hbar} \notag \\
r^j_{cv}(\mathbf{k}) &=& \frac{v^j_{cv}(\mathbf{k})}{i\omega_{cv}(\mathbf{k})} = \frac{\hbar v^j_{cv}(\mathbf{k})}{2 i d(\mathbf{k})} = \frac{-2B_j k_j \langle c | \Gamma_0 | v \rangle + A_j \langle c | \Gamma_j | v \rangle}{2 i d(\mathbf{k})}~,\notag
\end{eqnarray}
\noindent where in the last identity we have used $\langle c | I_{4} | v \rangle = 0$.

\subsection*{Choice of Tensor Elements}
The method of symmetry analysis of nonlinear optical conductivity tensors is well established \cite{boyd2008nonlinear}. 
The surface point group of 1T PtSe$_2$ is $C_{3v}$, (three fold rotation about the $z$-axis with a single mirror axis). The polycrystalline $PtSe_2$ in this paper consists of randomly oriented crystallites, further restricting the point surface group to $C_{\infty v}$, i.e. complete rotational symmetry. Under the action of $C_{\infty v}$, the only free non-zero elements of the PDE-tensor $\sigma^{ijkl}$ are: $(xxyy)$, $(xxzz)$, $(xyxy)$, $(xyyx)$, $(xzxz)$, $(xzzx)$, $(zxxz)$, $(zxzx)$, $(zzxx)$, $(zzzz)$, and the restriction $(xxx) = (xyyx) + (xxyy) + (xyxy)$. The photon drag contribution to the photocurrent is given by:
\begin{equation}
j^i_{\text{PD}} = \sigma^{jikl} q_{j} E_{k} E_{l}~,
\end{equation}
\noindent where for our experimental geometry consists of contacts lying on the $x$-axis, with light shone along the $y-z$ plane at $45^\circ$ angle of incidence. In this geometry the electric field of right/left circularly polarized light is given by $E_{R/L} = \left(\pm\dfrac{i}{\sqrt{2}},-\dfrac{1}{2},\dfrac{1}{2} \right)$, and the photon momentum is given by $\mathbf{q} = \dfrac{q}{\sqrt{2}}(0,1,1)$. The circular coefficient contribution to our measurement can be computed as the difference between the right and left circular responses: 
\begin{equation}
    \frac{1}{2}\left( j^x_{\text{PD}}(E_R) - j^x_{\text{PD}}(E_L) \right) = -\frac{q}{2} \text{Im} \left( \sigma^{xyxy}+\sigma^{zxxz} \right) 
\end{equation}

\subsection*{Strain Implementation}
To investigate the effect of mechanical deformation on the shift current, strain is introduced into the 3D BHZ model via a deformation potential approach. For a given uniform strain $\epsilon$, the primary modification occurs in the band inversion parameter $M$. The strained mass is modeled as:
\begin{equation}
M(\epsilon) = M_0 + \Lambda \epsilon
\end{equation}
\noindent where $M_0$ is the unstrained mass gap parameter and $\Lambda$ is the effective deformation potential (in eV per unit strain). This phenomenological modification captures the zeroth order band gap alteration one would expect from strains in most semiconductor systems [cite Pikus and Bir] as a function of the applied strain. While strain can also theoretically introduce anisotropic scaling to the Fermi velocities ($A_x, A_y, A_z$) due to lattice distortions and the Poisson effect, the current numerical implementation purely focuses on the lowest order impact of strain to the Hamiltonian, as deformation potentials are unknown for PtSe$_2$. Proper inclusion of deformation potentials would alter the evolution of the CPDE curves presented in the main text to be less monotonic, which would be more consistent with the experimentally observed trends. 

\subsection*{Comparing with CPGE}

For a thin film of PtSe$_2$ with multiple grain alignments and differences between substrate and atmosphere, it is natural to expect some structural inversion symmetry breaking to be present in the sample. In that case, it is also possible the photogalvanic (PGE) is responsible for the observed nonlinear photocurrent. We can augment the spinful BHZ Hamiltonian to include a Rashba spin-orbit coupling (SOC) gap and momentum term. At the zone center, the Rashba Hamiltonian takes the form $\alpha (\mathbf{k} \times \mathbf{\sigma}) \cdot \hat{z}$ for an inversion breaking field along (001). Expanding this near $K$ and $K'$ we find:
\begin{eqnarray}
H_{K/K',SOC}(\mathbf{k})
= H_{K/K'}(\mathbf{k}) +  \lambda  \alpha_0 \sigma_y \otimes \tau_0+ \alpha_R (k_x \sigma_y - k_y \sigma_x ) \otimes \tau_0
\end{eqnarray}
\noindent where $\lambda = \pm 1$ for $K/K'$. In principle the Rashba parameters $\alpha_0 \neq \alpha_R$, but for simplicity in our computations we assume equality. We can then compute the PGE contribution to the helical photocurrent as $\text{Im} (\sigma^{xxz})$ using the injection PGE formula:
\begin{eqnarray}
\sigma^{ijk}(\omega) &=&  \frac{i e^3/\hbar^2}{(\omega + i\delta)(-\omega + i\delta)}\times 
\notag\\
&~& \int d k \sum_{lmn} \frac{v_{nl}^i(\mathbf{k}) v_{lm}^j(\mathbf{k}) v_{mn}^k(\mathbf{k})}{ 2 i \delta - \epsilon_n + \epsilon_l }   \left[
\frac{f_{lm}}{\omega + i\delta +\epsilon_l - \epsilon_m }
-
\frac{f_{mn}}{-\omega + i \delta + \epsilon_m - \epsilon_n}
\right]~.\notag\\
\label{sigmadirty}
\end{eqnarray}
\noindent We perform the calculation for both $K$ and $K'$ Hamiltonians and sum them. The relevant tensors for CPGE are $\sigma^{xxz}$ and $\sigma^{xzx}$.



\begin{figure} 
	\centering
	\includegraphics[width=0.9\textwidth]{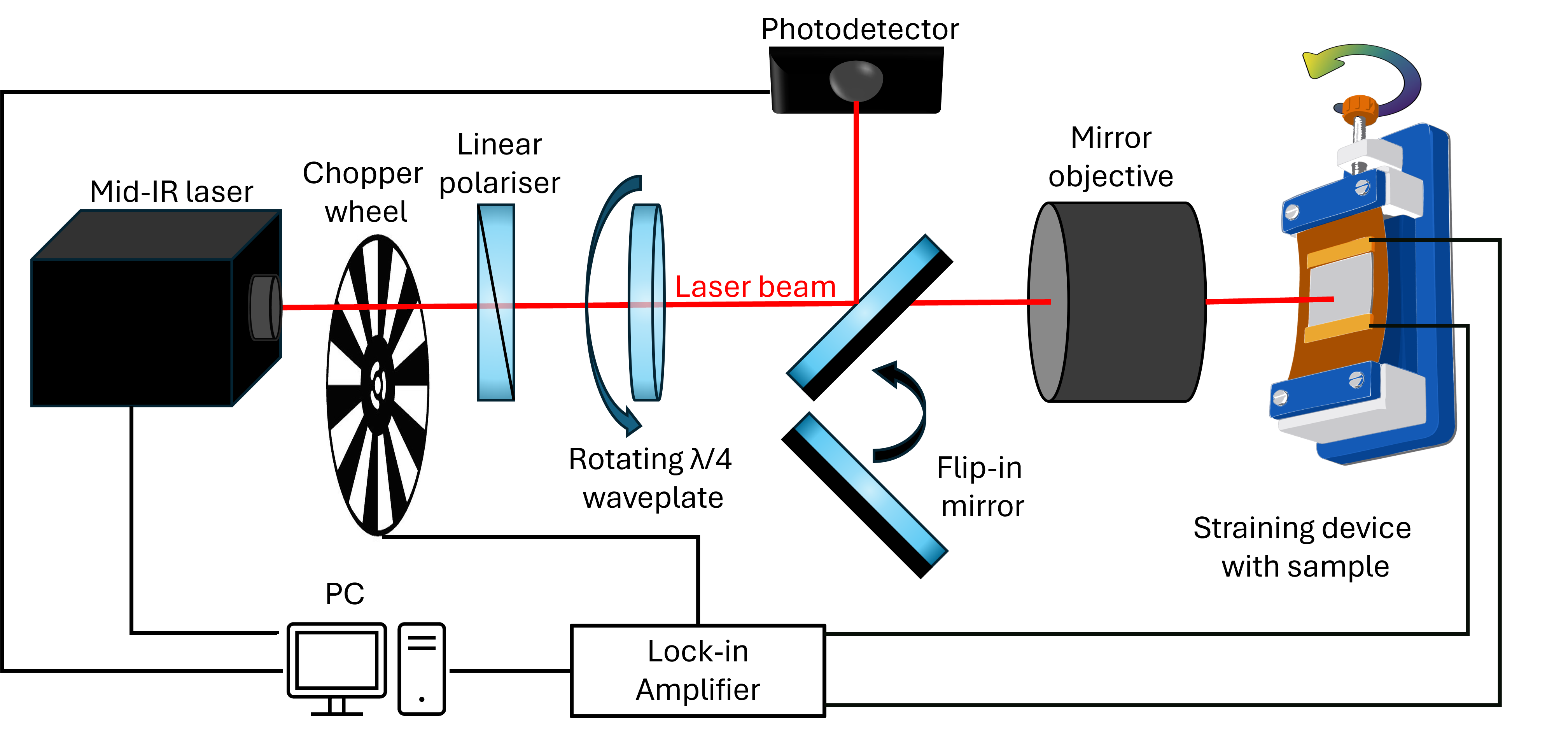} 

	\caption{Schematic of the mid-IR optical spectroscopy setup. A mid-IR QCL Laser source provides excitation with variable wavelengths ranging from $6 \upmu m $ to $11 \upmu m $. A chopper wheel was used to modulate the incoming signal and give a reference to the Lock-in Amplifier. A linear polarizer was used to set the linear polarization of the beam. The linearly polarized beam then passed a rotating quarter waveplate to create circularly polarized light. A flip-in mirror was used to measure the power of the laser beam with a photodetector. The beam was focused on the sample by a mirror objective to achieve similar focusing on a wide range of wavelengths. The photocurrent was measured using a lock-in amplifier and the measurement was controlled by a PC.} 
\label{fig:SI1}
\end{figure}

\begin{figure}
	\centering
	\includegraphics[width=0.9\textwidth]{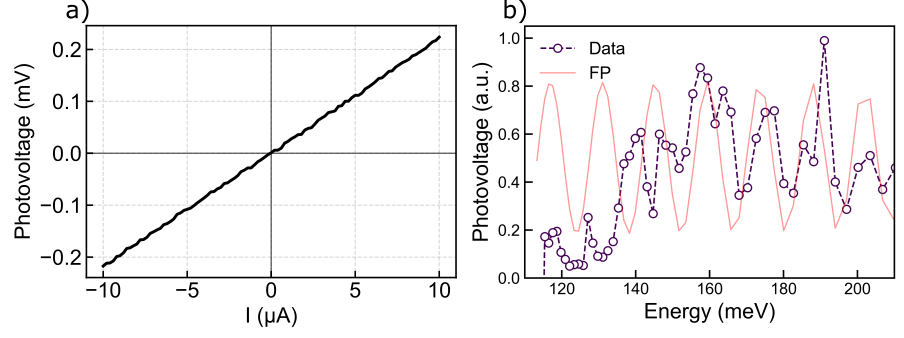} 
	\caption{a) The photovoltage of the PtSe$_2$ channel changes linearly under bias current. b) Fabry-Perot interference pattern (red) in the polyimid foil shows constructive and destructive interference with a period of 14.17 meV, approximated using an effective refractive index of 1.75 and thickness of 25 $ \upmu m$ for the polyimid foil. The data (purple) corresponds to the photoconductivity spectrum recorded at +10 $\upmu$A bias current in the unstrained state of the sample.}
\label{fig:SI2}
\end{figure}

\begin{figure}
	\centering
	\includegraphics[width=0.9\textwidth]{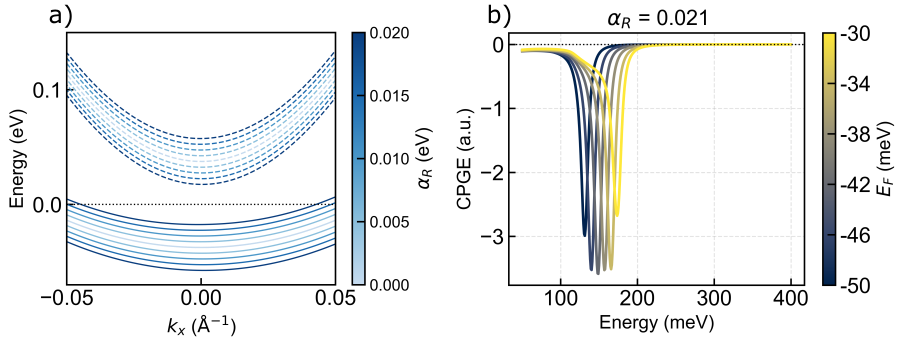} 
	\caption{a) Calculated Rashba splitting of the energy bands at the $K$-point for various $\alpha_R$ inversion breaking fields. b) Calculated CPGE spectrum depicted for a range of chemical potentials. The spectrum shows unipolar behavior and is shifting according to the chemical potential.}
\label{fig:SI3}
\end{figure}

\begin{figure}
	\centering
	\includegraphics[width=0.9\textwidth]{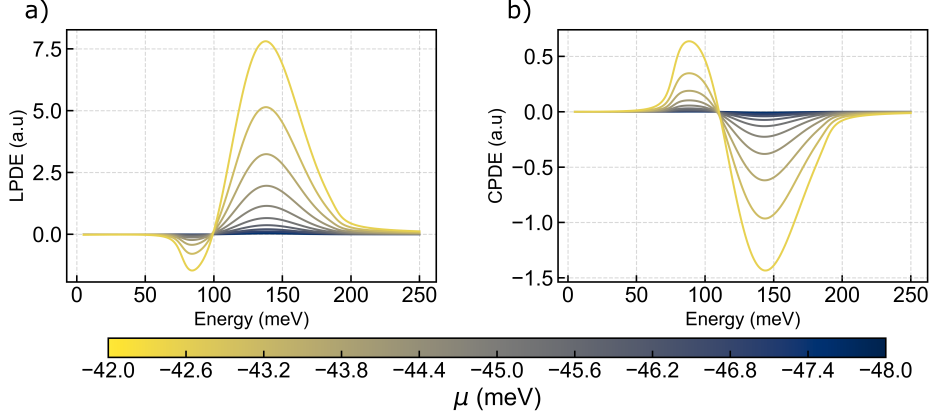} 
	\caption{Calculated spectrum of the linear photon drag effect a) and circular photon drag effect b) for various chemical potentials $\mu$. The shape of the calculated LPDE spectrum resembles the shape of the measured $L_1$ coefficient spectrum.} 
\label{fig:SI4}
\end{figure}

\begin{figure}
	\centering
	\includegraphics[width=0.9\textwidth]{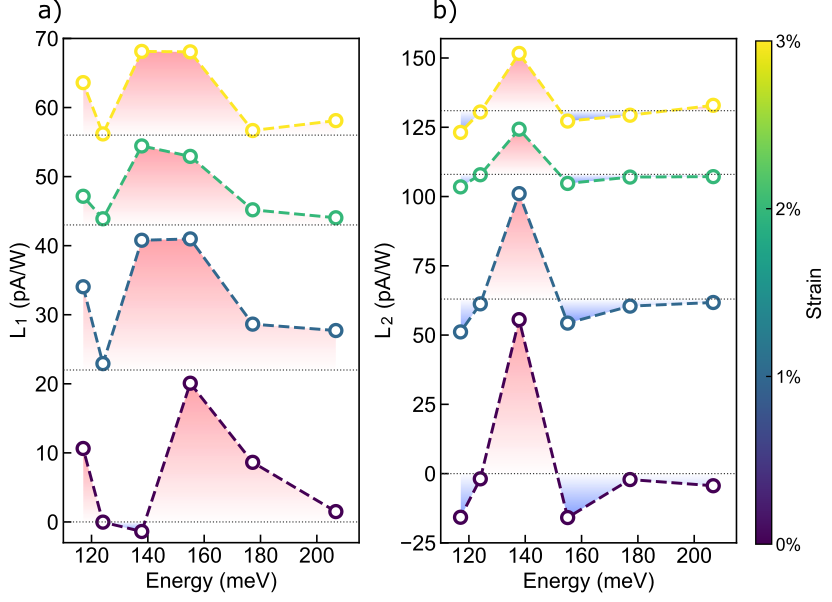} 
	\caption{a),b) Linear photocurrent $L_1$ and $L_2$ as a function of photon energy and for increasing uniaxial tensile strain. The curves for different strain values are offset for clarity. } 
\label{fig:SI5}
\end{figure}

\end{document}